\documentclass[aps,prd,reprint,showpacs,nofootinbib,longbibliography,preprintnumbers]{revtex4-1}
\usepackage{amsmath,amssymb,amsbsy,graphicx,color,xcolor,epsfig}
\usepackage{epstopdf}
\usepackage{hyperref}

\definecolor{darkred}{rgb}{0.75,0.0,0.0}
\definecolor{darkgreen}{rgb}{0.0,0.6,0.0}
\definecolor{darkblue}{rgb}{0.0,0.0,0.6}

\begin{document}

\preprint{
DESY 19-164\\
}

\title{ 
Interpretation of $Y_b (10750)$ as a tetraquark and its production mechanism
}

\author{Ahmed Ali} 
\email[]{Corresponding Author, Email: ahmed.ali@desy.de}
\affiliation{
Deutsches Elektronen-Synchrotron DESY, D-22607 Hamburg, Germany
}
\author{Luciano Maiani} 
\email[]{Corresponding Author, Email: luciano.maiani@cern.ch}
\affiliation{
T.\,D.~Lee Institute, Shanghai Jiao Tong University, Shanghai, 200240, China
}
\affiliation{
Dipartimento di Fisica and INFN, Sapienza Universit\`{a} di Roma, Piazzale Aldo Moro 2, I-00185 Roma, Italy
} 
\author{Alexander Ya. Parkhomenko} 
\email[]{Corresponding Author, Email: parkh@uniyar.ac.ru}
\affiliation{
Department of Theoretical Physics, P.\,G.~Demidov Yaroslavl State University, 
Sovietskaya 14, 150003 Yaroslavl, Russia
}
\author{Wei Wang} 
\email[]{Corresponding Author, Email: wei.wang@sjtu.edu.cn}
\affiliation{
INPAC, SKLPPC, MOE KLPPC,
School of Physics and Astronomy, Shanghai Jiao Tong University, Shanghai 200240, China 
}

\begin{abstract}
Recently, the Belle Collaboration has updated the analysis of the cross sections for the processes
$e^+ e^- \to \Upsilon(nS)\, \pi^+ \pi^-$ ($n = 1,\, 2,\, 3$) in the $e^+ e^-$ center-of-mass energy
range from 10.52 to 11.02~GeV. A new structure, called 
here $Y_b (10750)$, with the mass $M (Y_b) = (10752.7 \pm 5.9^{+0.7}_{-1.1})$~MeV and the Breit-Wigner 
width $\Gamma (Y_b) = (35.5^{+17.6 +3.9}_{-11.3 -3.3})$~MeV was observed~\cite{Abdesselam:2019gth}. 
We interpret $Y_b (10750)$ as a compact $J^{PC} = 1^{--}$ state with a dominant tetraquark component.
The mass eigenstate $Y_b (10750)$ is treated as a linear combination of the diquark-antidiquark 
and $b \bar b$ components due to the mixing via gluonic exchanges shown recently to arise in the limit 
of large number of quark colors. The mixing angle between $Y_b$ and $\Upsilon(5S)$ can be estimated  
from the electronic width, recently determined to be $\Gamma_{ee} (Y_b) = (13.7 \pm 1.8)$~eV.  
The mixing provides a plausible mechanism for $Y_b (10750)$ production in high energy collisions 
from its $b \bar b$ component and we work out the Drell-Yan and prompt production 
cross sections for $p p \to Y_b (10750) \to \Upsilon (nS)\, \pi^+ \pi^-$ at the LHC.  
The resonant part of the dipion invariant mass spectrum in $Y_b (10750) \to \Upsilon (1S)\, \pi^+ \pi^-$ and 
the corresponding angular distribution of $\pi^+$-meson in the dipion rest frame are presented as an example. 
\end{abstract}

\date{\today}

\maketitle 

%

\leftline{\bf 1. Introduction}   
\vspace*{\baselineskip}

Recently, Belle has reported an updated measurement of the cross sections for 
$e^+ e^- \to \Upsilon (nS)\, \pi^+ \pi^-$; $nS = 1S,\, 2S,\, 3S$ in the $e^+ e^-$ center-of-mass 
energy range from 10.52 to 11.02~GeV. 
They observe a new structure, $Y_b (10750)$, in addition to the $\Upsilon (10860)$- and 
$\Upsilon (11020)$-resonances, having the masses and Breit-Wigner decay widths shown 
in Table~\ref{tab:Belle-data-1-2019}~\cite{Abdesselam:2019gth}.
The measured ranges of the product $\Gamma_{ee} \times {\cal B}$ (in~eV) for the three final states 
are also presented in Table~\ref{tab:Belle-data-1-2019}. The global significance of the new structure 
is~5.2$\sigma$. 
We recall that in high statistics energy scans for the ratios 
$R_{\Upsilon\, \pi^+ \pi^-} \equiv 
\sigma (e^+ e^- \to (\Upsilon (1S),\, \Upsilon (2S),\, \Upsilon (3S))\, \pi^+ \pi^-)/
\sigma (e^+ e^- \to \mu^+ \mu^-)$ and $R_{b \bar b} \equiv 
\sigma (e^+ e^- \to b \bar b)/\sigma(e^+ e^- \to \mu^+ \mu^-)$, 
Belle had found no new structures in their 2016 analysis~\cite{Santel:2015qga}.
In the same analysis, a 90\%~C.L. upper limit of 9~eV was set on~$\Gamma_{ee}$ 
in search of a structure around 10.9~GeV in $R_{b \bar b}$~\cite{Santel:2015qga}.
We also recall that the visible cross section for $e^+ e^- \to B_s^{(*)} \bar{B}_s^{(*)}$ showed 
a clear peak for $\Upsilon (10860)$, a less clear one for the $\Upsilon (11020)$, but no significant 
signal was observed around 10.75~GeV~\cite{Abdesselam:2016tbc}. 

The combined BaBar and Belle data on $R_{b \bar b}$ have been recently reanalyzed taking into account 
the coherent sum of the three resonances $\Upsilon (10860)$, $\Upsilon (11020)$, and $Y_b (10750)$~\cite{CZY}, 
and a continuum amplitude, proportional to $1/\sqrt s$, where $\sqrt s$ is the center-of-mass $e^+ e^-$ energy. 
The fit parameters of the $R_{b \bar b}$ lineshape are the masses, Breit-Wigner decay widths, leptonic partial 
decay widths, and the relative phases. The resulting resonance masses and decay widths are found to be 
in agreement with the ones obtained from the $R_{\Upsilon\, \pi^+ \pi^-}$ scan, and one gets a number 
of solutions for the partial electronic widths (mathematically 8~solutions are expected), which differ 
in other parameters, such as the total decay widths and partial leptonic widths. Most of these solutions are 
likely unphysical except the first solution, in which the electronic width of~$Y_b$ is given as~\cite{CZY}: 
\begin{eqnarray} 
\Gamma_{ee} (Y_b (10750)) &=& \left ( 13.7 \pm 1.8 \right ) {\rm eV}. \label{mixval}
\end{eqnarray}

\begin{figure}
 \begin{center}
   \includegraphics[width=0.8\columnwidth]{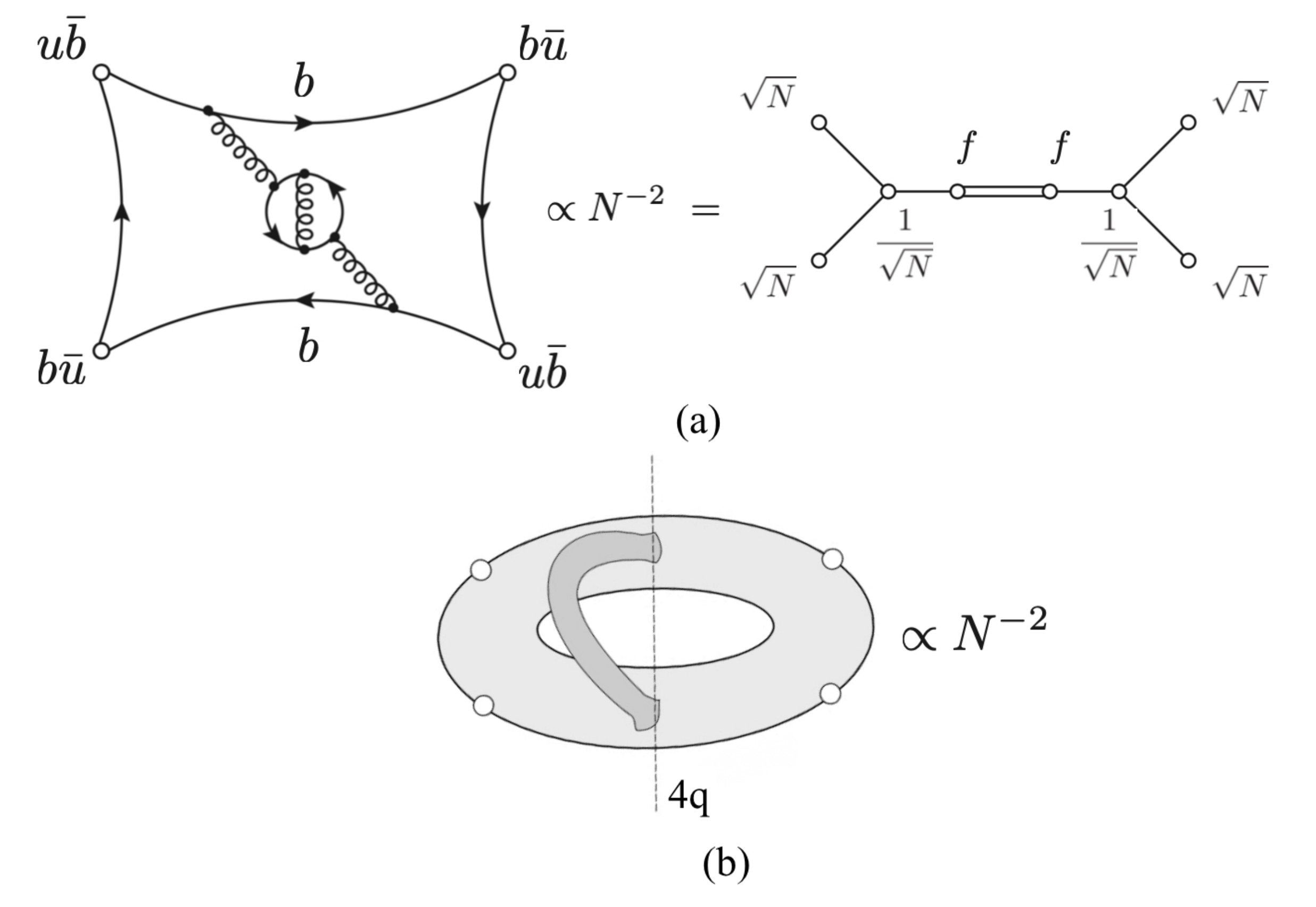}
 \end{center}
\caption{\footnotesize 
(a) Left-hand side: lowest order diagram for meson-meson scattering that may have quarkonium 
and four-quark poles connected by mixing, as indicated by the diagram on the right-hand side, 
see Eq.~(\ref{eq:orderf}).  
(b) Topological structure of the nonperturbative realization of the same process. 
$N$~denotes the number of colors.
}  
\label{mixingdiag}
\end{figure}

In this paper, we interpret $Y_b (10750)$ as a $J^{PC} = 1^{--}$ tetraquark candidate, 
whose dominant component~$Y_b^0$ consists of a colored diquark-antidiquark pair 
$[b q]_{\bar 3_c} [\bar b \bar q]_{3_c}$, bound in the $SU (3)$ antitriplet-triplet 
representation~\cite{Jaffe:2003sg,Maiani:2004vq}. However, it can have a small $b \bar b$ 
component due to the mixing via gluonic exchanges. 
The behavior of QCD for large-$N_c$, where~$N_c$ is the number of colors,
has been worked out long ago by 't~Hooft~\cite{tHooft:1973alw}. With the quark-gluon coupling 
as ${\it L}_{\rm QCD} = g_{\rm QCD} \bar q \lambda^A g_\mu^A \gamma^\mu q$ and $\lambda^A$ 
being the $N_c^2 - 1$ $SU (N_c)$ matrices, the large-$N_c$ limit is considered 
as $g_{\rm QCD} \to 0$, $g_{\rm QCD}^2 N_c \equiv \lambda$ fixed. 
The amplitudes are expanded in powers of $1/N_c$, with each term being a nonperturbative function 
of the reduced coupling~$\lambda$. As explained below (see Fig.~\ref{mixingdiag}), it implies that 
$\Upsilon (10860)$ and $\Upsilon (11020)$, which are dominantly radial $b \bar b$ excitations, 
$\Upsilon (5S)$ and $\Upsilon (6S)$, respectively, also have a small diquark-antidiquark 
component~$Y_b^0$ in their Fock space. Due to the proximity of the mass eigenstates $Y_b (10750)$ 
and $\Upsilon (10860)$, we consider that the mixing is dominantly between~$Y_b^0$ and $\Upsilon (5S)$. 
This also provides a plausible interpretation of some anomalous features measured in the decays 
of the $\Upsilon (10860)$.\footnote{
A tetraquark interpretation~\cite{Ali:2009es,Ali:2010pq} had been put forward for the $Y_b (10890)$, 
a resonance observed by Belle more than a decade ago~\cite{Abe:2007tk,Adachi:2008pu}, together with 
$Y_b (10860)$, identified with $\Upsilon (5S)$. In subsequent data by Belle~\cite{Santel:2015qga}, 
two states $Y_b (10890)$ and $\Upsilon (10860)$ were found to have the same mass within~2$\sigma$, 
essentially closing the window for an additional resonance. This seems to have changed with 
the announcement of $Y_b (10750)$.}

We argue that the production mechanism of $Y_b (10750)$ proceeds through its $b \bar b$ component, 
which arises from the mixing $([b q]_{\bar 3_c} [\bar b \bar q]_{3_c} - b \bar b)$.    
A non-vanishing mixing is induced by non-planar diagrams~\cite{Maiani:2018pef}, allowing 
the direct production of $Y_b (10750)$ in high energy collisions.
Using this, Drell-Yan~\cite{Ali:2011qi} and prompt production cross sections~\cite{Ali:2013xba} 
for $Y_b (10750)$ are presented for the LHC. We estimate the $Y_b - \Upsilon (5S)$ 
mixing angle from $\Gamma_{ee} (Y_b)$ in Eq.~\eqref{mixval}

\begin{table}[tb]
\caption{
Measured masses and decay widths (in MeV), and ranges of $\Gamma_{ee} \times {\cal B}$ (in eV) of the $\Upsilon(10860)$, $\Upsilon(11020)$, and the new structure $Y_b(10750)$. The first uncertainty is
statistical and the second is systematic (Belle~\cite{Abdesselam:2019gth}).
}
\label{tab:Belle-data-1-2019} 
\begin{center}
\hspace{-4mm}
\begin{tabular}{cccc} 
\hline
State        & $\Upsilon (10860)$    &     $\Upsilon (11020)$     &     $Y_b(10750)$     \\ \hline      
Mass  & {\footnotesize $10885.3 \pm 1.5^{+2.2}_{-0.9}$} & {\footnotesize $11000.0^{+4.0  +1.0}_{-4.5 -1.3}$} & {\footnotesize  $10752.7 \pm 5.9^{+0.7}_{-1.1}$} \\
Width & {\footnotesize  $36.6^{+4.5 +0.5}_{-3.9 -1.1}$} & {\footnotesize     $23.8^{+8.0 +0.7}_{-6.8 -1.8}$} & {\footnotesize $35.5^{+17.6 +3.9}_{-11.3 -3.3}$} \\
\hline  
$\Upsilon(1S) \pi^+ \pi^-$ & {\footnotesize $0.75 - 1.43$} & {\footnotesize $0.38 - 0.54$} & {\footnotesize $0.12 - 0.47$} \\[1mm] 
$\Upsilon(2S) \pi^+ \pi^-$ & {\footnotesize $1.35 - 3.80$} & {\footnotesize $0.13 - 1.16$} & {\footnotesize $0.53 - 1.22$} \\[1mm] 
$\Upsilon(3S) \pi^+ \pi^-$ & {\footnotesize $0.43 - 1.03$} & {\footnotesize $0.17 - 0.49$} & {\footnotesize $0.21 - 0.26$} \\[1mm] \hline 
\end{tabular}
\end{center} 
\end{table}

In contrast to the decays of $\Upsilon (10860)$ and $\Upsilon (11020)$, whose dipionic transitions 
$(\Upsilon (1S),\, \Upsilon (2S),\, \Upsilon (3S))\, \pi^+ \pi^-$ are dominated by the resonant 
$Z^\pm_b (10650)$ and $Z^\pm_b (10610)$ states~\cite{Belle:2011aa}, the decay 
$Y_b (10750) \to Z^\pm_b (10650)\, \pi^\mp$ is kinematically forbidden, and
$Y_b (10750) \to Z^\pm_b (10610)\, \pi^\mp$ has a strong phase-space suppression. 
Thus, $Y_b (10750)$ decays are anticipated to reflect their dominant non-resonant component. 
In addition, the decays $Y_b \to (\Upsilon (1S),\, \Upsilon (2S),\, \Upsilon (3S))\, \pi^+ \pi^-$,
being Zweig-allowed, are anticipated to have decay widths characteristic of strong interactions. 
Dalitz analysis in the decay $Y_b \to \Upsilon (1S)\, \pi^+ \pi^-$ will show a band structure 
in the $m_{\pi^+ \pi^-}$ invariant mass, revealing clear evidence of two scalars, $f_0 (500)$ 
and $f_0 (980)$, and the tensor $J^{PC} = 2^{++}$ meson, $f_2 (1270)$~\cite{Tanabashi:2018oca}. 
In other two decays 
$Y_b \to (\Upsilon (2S),\, \Upsilon (3S))\, \pi^+ \pi^-$, only the broad $f_0 (500)$-meson is present. 
With higher statistics data anticipated with the Belle-II detector, this distribution, as well as 
other properties of $Y_b (10750)$, will be well measured, allowing us to discriminate the tetraquark 
picture from other competing mechanisms, such as a $D$-wave interpretation of $Y_b (10750)$, with 
a large $S - D$ mixing~\cite{Badalian:2009bu}.


\vspace*{\baselineskip}
\leftline{\bf 2. Tetraquark-$Q \bar Q$ mixing in large-$N_c$ approach}   
\vspace*{\baselineskip}

In a seminal paper, S.~Weinberg~\cite{Weinberg:2013cfa} addressed the description of tetraquarks 
in the large-$N_c$ limit, followed by several investigations~\cite{%
Knecht:2013yqa,Lebed:2013aka,Cohen:2014tga,Maiani:2016hxw,Lucha:2017mof,Lucha:2017gqq,Maiani:2018pef}.  
A mixing between a bottomonium and hidden-beauty tetraquark, anticipated in~\cite{Knecht:2013yqa}, was shown 
in~\cite{Maiani:2018pef}, to be induced at the level of non-planar diagrams, Fig.~\ref{mixingdiag}(a) and~(b).

In brief, the exchange of a gluon between the two quark loops in Fig.~\ref{mixingdiag}(a) produces 
the interaction by which a genuine tetraquark pole may form in the intermediate state. 
Fig.~\ref{mixingdiag}(b) displays the non-perturbative version of Fig.~\ref{mixingdiag}(a). 
In the language introduced by 't~Hooft for the large-$N_c$ expansion~\cite{tHooft:1973alw}, 
non-planar gluon exchanges between the two fermion loops mean topologically one handle 
and produce a mixing coefficient~$f$ of order\footnote{
In the large-$N_c$ language, an amplitude ${\cal A}$ for a process has the dependence 
${\cal A} \propto N_c^\alpha$, where~$\alpha = 2 - L - 2 H$, with~$L$ 
being the number of fermion loops and~$H$ the number of handles, i.\,e. independent 
non-planar sets of gluons. For a planar diagram $H = 0$ and $L = 1$, yielding $\alpha = 1$.
Large-$N_c$-counting rules in the context of tetraquarks are given in~\cite{Ali:2019roi}.
}
\begin{equation}
f = \frac{1}{N_c \sqrt{N_c}} . 
\label{eq:orderf}
\end{equation}
 
A non-vanishing mixing with charmonia is also predicted in the alternative extension 
to large~$N_c$ based on Witten's picture of large-$N_c$ baryons~\cite{Witten:1979kh}. 
These ``generalized tetraquarks'' are made by $N_c - 1$ antisymmetric quarks bound 
to $N_c - 1$ antisymmetric antiquarks~\cite{Rossi:1977cy,Montanet:1980te,Cohen:2014vta,Rossi:2016szw}. 
Non-vanishing mixing with quarkonia has been noted in~\cite{Maiani:2018tfe}. 
Albeit suppressed at large~$N_c$ by the exponential factor $e^{-N_c/2}$, when extrapolated 
back to $N_c = 3$ one finds a result not dissimilar from~\eqref{eq:orderf}. Thus, production 
in the $e^+ e^-$-annihilation of resonances such as $Y_b (10750)$, in addition to the bottomonia 
spectral lines and with a small $\Gamma_{ee}$, is a significant signature of tetraquarks.


\vspace*{\baselineskip}
\leftline{\bf 3. Mixing formalism}   
\vspace*{\baselineskip}

Following~\cite{Ali:2009es,Ali:2010pq}, we define the tetraquark states~$Y_b^I$ 
in the isospin basis, with the  
$Y_b^0 \equiv \left ( Y_{[bu]} + Y_{[bd]} \right )/\sqrt 2$ and
$Y_b^1 \equiv \left ( Y_{[bu]} - Y_{[bd]} \right )/\sqrt 2$
for isospin $I = 0$ and $I = 1$, respectively. We ignore their mass difference due 
to the isospin breaking. Since the production is via the isosinglet $b \bar b$-component, 
we consider only~$Y_b^0$, the isospin-0 state. In view of the observed mass difference 
(see Table~\ref{tab:Belle-data-1-2019}) $M [\Upsilon (10860)] - M [Y_b (10750)] \simeq 133$~MeV, 
compared to the mass difference $M [\Upsilon (11020)] - M [Y_b (10750)] \simeq 247$~MeV, 
we only consider the mixing between $\Upsilon (10860)$ and $Y_b (10750)$, though it 
can be generalized to the case with all three states.

Mass eigenstates are rotated from the eigenstates in the quark flavor space, 
with the latter defined as $\Upsilon (5S)$ and~$Y_b^0$, respectively.
\begin{equation}
\left (
\begin{array}{c} 
      Y_b (10750) \\
 \Upsilon (10860)
\end{array}
\right ) 
=
\left ( 
\begin{array}{rr} 
  \cos\tilde\theta & \sin\tilde\theta \\
- \sin\tilde\theta & \cos\tilde\theta
\end{array}
\right ) 
\left (
\begin{array}{c} 
         Y_b^0 \\
 \Upsilon (5S)
\end{array}
\right ) ,
\label{eq:state-rotation}
\end{equation}
where $\tilde\theta$ is a mixing angle, estimated below phenomenologically.
This mixing relates $\Gamma_{ee} [Y_b (10750)]$ and $\Gamma_{ee} [\Upsilon (5S)]$, yielding 
\begin{equation}
\frac{\Gamma_{ee} [Y_b (10750)]}{\Gamma_{ee} [\Upsilon (10860)]} = 
\tan^2\tilde\theta\,   \left [ 
\frac{M [\Upsilon (10860)]}{M [Y_b (10750)]} 
\right ]^4 \simeq 1.04\, \tan^2\tilde\theta\,  . 
\label{eq:Gamma-ee-Yb-5S}
\end{equation}
Recalling that $\Gamma_{ee} [\Upsilon (10860)] = (310 \pm 70)$~eV~\cite{Tanabashi:2018oca}, 
and the recent value given in~\eqref{mixval}, we find
\begin{equation}
\tan^2\tilde\theta = 0.044 \pm 0.006,
\label{eq:kappa-bound}
\end{equation}
 which leads to $\tilde\theta \sim 12^\circ$.


\begin{flushleft}
{\bf 4. Hadroproduction and Drell-Yan cross sections for $pp \to Y_b (10750) \to \Upsilon (nS)\, \pi^+ \pi^-$ at the LHC}   
\end{flushleft}

%
In Ref.~\cite{Ali:2013xba}, the hadroproduction cross sections for $\Upsilon (5S)$ and $\Upsilon (6S)$ in $p \bar p$ 
and $p p$ collisions were calculated at the Tevatron and LHC, using the Non-Relativistic QCD framework~\cite{Bodwin:1994jh}.  
The calculation has adopted a factorization ansatz to separate the short- and long-distance effects.

First, cross-sections for $Y_b (10750)$ are scaled from the ones for $\Upsilon (5S)$, since the production 
takes place via the $b \bar b$-component in the $Y_b (10750)$ Fock space. The latter is determined by 
the mixing angle, derived in the previous section, and results in the following relation:
\begin{eqnarray}
\frac{\sigma ( p p \to Y_b (10750) + X)\, {\cal B}_f (Y_b)}
     {\sigma ( p p \to \Upsilon (10860) + X)\, {\cal B}_f (\Upsilon (10860))}
\nonumber \\
\hspace{-30mm}\simeq 
\frac{\Gamma_{ee} (Y_b)\, {\cal B}_f (Y_b)} 
     {\Gamma_{ee} (\Upsilon (10860))\,  {\cal B}_f (\Upsilon (10860))} .
\label{eq:Yb-Upsilon}
\end{eqnarray} 
Here, ${\cal B}_f (Y_b)$ and ${\cal B}_f (\Upsilon (10860))$ denote the branching ratios of
$Y_b(10750) \to f$ and $\Upsilon(10860) \to f$, respectively, where $f$ represents the three dipionic
final states $\Upsilon(1S) \pi^+ \pi^-$, $\Upsilon(2S) \pi^+ \pi^-$, and $\Upsilon(3S) \pi^+ \pi^-$.
The r.h.s. of Eq.~(\ref{eq:Yb-Upsilon}) has been measured by Belle~\cite{Abdesselam:2019gth}.

Secondly, to obtain the absolute cross section for $Y_b (10750)$ production, we estimate 
the $\Upsilon (10860)$ cross section in NRQCD, following the calculation presented in~\cite{Ali:2013xba}.
One starts from the formula:
\begin{eqnarray}
&& \sigma (p p \to \Upsilon (10860) + X) = \sum _Q \sigma_Q \notag \\
&& \hspace{7mm}
= \sum_Q \int dx_1 dx_2 \sum_{i,j} f_i (x_1) f_j (x_2) \notag \\  
&& \hspace{15mm} 
\times \hat\sigma \left ( i j \to \langle \bar b b \rangle_Q + X \right ) \langle O [Q] \rangle . 
\end{eqnarray}
Here,~$i$ and~$j$ denote a generic parton inside a proton, 
$f_i (x_1)$ and $f_j (x_2)$ are the parton distribution functions (PDFs)~\cite{Nadolsky:2008zw}, 
the label~$Q$ denotes the quantum numbers of the $b \bar b$-pair, which are labeled as 
$^{2S+1} L_J^c$ (color~$c$, spin~$S$, orbital angular momentum~$L$ and total angular momentum~$J$),  
$\langle O [Q] \rangle$ are the corresponding long-distance matrix elements (LDMEs), 
and~$\hat\sigma$ is a partonic cross section.

The leading-order partonic processes for the $S$-wave configurations are:
\begin{align} 
g + g 
&\to
\Upsilon [^3 S_1^1] + g,
\nonumber\\
g  +  g 
&\to
\Upsilon [^1 S_0^8,\; ^3 S_1^8]  + g, 
\nonumber\\
g + q 
&\to
\Upsilon [^1 S_0^8,\; ^3 S_1^8]  + q, 
\nonumber\\
q  + \bar q 
&\to
\Upsilon [^1 S_0^8,\; ^3 S_1^8]  + g . 
\end{align}
The normalized cross sections, in which the LDMEs are factored out, are defined by 
$\tilde\sigma_Q \equiv \sigma_Q/\langle O [Q] \rangle$. They have been calculated 
in Ref.~\cite{Ali:2013xba} for the LHC energies $\sqrt s = 7$, 8 and~14~TeV. 
They are supplemented by the LDMEs, for which the following values have been used. 
The LDME of the Color-Singlet $^3 S^1_1 $ is $\langle  O [Q]\rangle \simeq 0.56$~GeV$^3$; 
the Color-Octet LDMEs, for $^1 S^8_0$ and $^3 S^8_1$, are estimated as 
$\langle O [Q] \rangle = (-0.95 \pm 0.38) \times 10^{-3}$ GeV$^3$, 
and $\langle O [Q] \rangle = (3.46 \pm 0.21) \times 10^{-2}$ GeV$^3$, respectively.  
Summing over the partonic processes shown above, and using the branching ratios from the PDG, yields 
the cross sections $\sigma (p p \to \Upsilon (5S) \to (\Upsilon (nS) \to \mu^+ \mu^-)\, \pi^+ \pi^-)$, 
where $n = 1,\, 2,\, 3$~\cite{Ali:2013xba}. 
 
\begin{table*}
\caption{
Total cross sections (in~pb) for the processes
$p p \to Y_b (10750) \to (\Upsilon(nS) \to \mu^+\mu^-)\, \pi^+\pi^-$ $(n = 1,\, 2,\, 3)$  
at the LHC ($\sqrt s = 14$~TeV), assuming the transverse momentum range $3~{\rm GeV} < p_T < 50~{\rm GeV}$.  
The rapidity range $|y| < 2.5$ is used for ATLAS and CMS (called LHC~14), and the rapidity range $2.0 < y < 4.5$ is used 
for the LHCb. The error estimates in the QCD production are from the variation of the central values 
of the Color-Octet LDMEs and the various decay branching ratios, as discussed in Ref.~\cite{Ali:2013xba}.  
Contributions from $\Upsilon (1S,\, 2S,\, 3S)$ are added together in the Drell-Yan production mechanism 
as in Ref.~\cite{Ali:2011qi}.  
}
\label{tab:fullresult}
\begin{tabular}{lccc|cccc}
\hline\hline
 &
 &
 QCD (gg)
 &
 & 
 Drell-Yan
 &
 \\
 & $n = 1$ & $n = 2$ & $n = 3$
 & DY\\\hline 
LHC 14
 & [ 0.29,  3.85]
 & [ 0.70,  4.78]
 & [ 0.45,  3.10] 
 & [0.002, 0.004]
\\
LHCb 14
 & [ 0.08,  1.21]
 & [ 0.20,  1.51]
 & [ 0.13,  0.99] 
 & [0.001, 0.002]
\\
 \hline\hline
\end{tabular} 
\end{table*}

The corresponding cross sections for the processes 
$p p \to Y_b (10750) \to (\Upsilon(nS) \to \mu^+\mu^-)\, \pi^+\pi^-$
are obtained by using the scaling relation given in Eq.~(\ref{eq:Yb-Upsilon}).
For the LHC at $\sqrt s = 14$~TeV, cross sections are given in Table~\ref{tab:fullresult}  
for the indicated ranges of $p_T (Y_b)$ and rapidity~$|y|$, separately for ATLAS/CMS 
and for LHCb. Theoretical uncertainties in these cross sections are almost a factor~10, dominated by the 
uncertainties on the Color-Octet LDMEs, as well as on the ratio on the r.h.s. in Eq.~(\ref{eq:Yb-Upsilon}). 
To estimate the expected number of events, we use 1~pb for the cross section, which lies in the middle 
of the indicated ranges, yielding $O (10^4)$ signal events at the LHCb, and an order of magnitude larger 
for the other two experiments, ATLAS and CMS. The discovery channel $\mu^+ \mu^-\, \pi^+ \pi^-$,
with the $\mu^+ \mu^-$ mass constrained by the $\Upsilon (nS)$ ($nS = 1S,\, 2S,\, 3S$) masses, involves 
a pair of charged pions. Thus, the background is a stumbling block, but hopefully this can be overcome, 
with the additional constraint of the $Y_b (10750)$ mass. In addition to the mixing mechanism utilized here, 
there maybe direct production of the tetraquark, which would add incoherently to the previous results. 
Thus, the numbers presented in Table~\ref{tab:fullresult} give lower bounds to the expected $Y_b (10750)$ 
production in $pp$ collisions. 

The Drell-Yan production cross sections and differential distributions in the transverse momentum
and rapidity of the $J^{PC} = 1^{--}$ exotic hadrons $\phi (2170)$, $X (4260)$ and $Y_b (10890)$ 
at the hadron colliders LHC and Tevatron have been calculated in~\cite{Ali:2011qi}. We update these 
calculations for the production of $Y_b (10750)$ at the LHC for $\sqrt s = 14$~TeV, and present results 
for $p p \to Y_b (10750) \to (\Upsilon (nS) \to \mu^+ \mu^-)\, \pi^+ \pi^-$ taking into account 
the current mass of $Y_b (10750)$ and the measured quantity $\Gamma_{ee} \times {\cal B}$, whose 
ranges are measured by Belle~\cite{Abdesselam:2019gth} and given in Table~\ref{tab:Belle-data-1-2019}.
In deriving the distributions and cross sections, we have included the order~$\alpha_s$ QCD corrections, 
resummed the large logarithms in the small transverse momentum region in the impact-parameter formalism, 
and have used two sets of parton distribution functions: MSTW (Martin-Stirling-Thorne-Watt) 
PDFs~\cite{Martin:2009iq} and CTEQ10~\cite{Lai:2010vv}; the details can be seen in~\cite{Ali:2011qi}.
Numerical results for the cross section are given in Table~\ref{tab:fullresult}, where the~$p_T$ and 
rapidity~$|y|$ ranges for the ATLAS and CMS (called LHC~14 there), and for the LHCb, are indicated. 
These cross sections yield $O (300)$ events for the current ATLAS/CMS luminosity (140~fb$^{-1}$), 
and $O (10)$ events for the LHCb (9~fb$^{-1}$), but could be higher by a factor~2. The Drell-Yan 
cross sections are theoretically more accurate, but suffer from the small rates compared to the 
hadroproduction cross sections at the LHC.



\begin{flushleft}
{\bf 5. Dipion invariant mass spectra and angular distributions in $e^+ e^- \to Y_b \to \Upsilon (nS)\, \pi^+ \pi^-$}   
\end{flushleft}

%
The amplitudes of the $e^+ e^- \to Y_b \to \Upsilon (nS)\, P P'$ process, where $P^{(\prime)}$ 
is a pseudoscalar, have been calculated in~\cite{Ali:2010pq} as a sum of the Breit-Wigner resonances 
and non-resonating continuum contributions, with the latter adopted from~\cite{Brown:1975dz}. 
The differential cross section is then written as~\cite{Ali:2010pq}: 
\begin{align}
\frac{d^2\sigma_{\Upsilon (1S) P P'}}{dm_{P P'}\, d\cos\theta}
&=
\frac{
\lambda^{1/2} (s, m_\Upsilon^2, m_{P P'}^2)
\lambda^{1/2} (m_{P P'}^2, m_P^2, m_{P'}^2)
}{
384\pi^3 s\, m_{P P'}
\left [ (s - m_{Y_b}^2 )^2 + m_{Y_b}^2 \Gamma_{Y_b}^2 \right ] 
}
\nonumber\\
&\hspace{-18mm} \times
\Bigg \{
\left ( 1 + \frac{(q \cdot p)^2}{2 s\, m_\Upsilon^2} \right )
\left | {\cal S} \right |^2
\nonumber\\
&\hspace{-13mm}
+
2\, \text{Re} \left [ {\cal S}^*
\left ( {\cal D}' + \frac{(q\cdot p)^2}{2 s\, m_\Upsilon^2}\, {\cal D}'' \right )
\right ]
\left ( \cos^2\theta - \frac{1}{3} \right )
\nonumber\\
&\hspace{-13mm}
+
\left | {\cal D} \right |^2 \sin^2\theta
\left [ \sin^2\theta + 2 \left ( 
\frac{q_0^2}{s} + \frac{p_0^2}{m_\Upsilon^2} 
\right ) \cos^2\theta
\right ]
\nonumber\\
&\hspace{-13mm}
+
\left (
\left | {\cal D}' \right |^2 + \frac{(q \cdot p)^2}{2 s\, m_\Upsilon^2} \left | {\cal D}'' \right |^2
\right ) \left ( \cos^2\theta - \frac{1}{3} \right )^2
\Bigg \}\, ,
\end{align}
where $s$ and $m_{P P^\prime}$ are the squared invariant masses of the $e^+ e^-$-pair and 
a pair of two final pseudocsalars,~$\theta$ is the angle between the momenta of~$Y_b$ and~$P$ 
in the $P P^\prime$ rest frame, $\lambda (x, y, z) \equiv (x - y - z)^2 - 4 y z$,~$q_0$ and~$p_0$ 
are the energies of the $Y_b$- and $\Upsilon (1S)$-mesons in the $P P'$ rest frame, respectively, 
$\Gamma_{Y_b}$ is the decay width of~$Y_b$, and $m_{Y_b}$, $m_\Upsilon$, $m_P$ and $m_{P'}$ 
are the masses of $Y_b$, $\Upsilon (1S)$, $P$ and $P'$, respectively.

%
 
The $S$-wave amplitude for the $P P'$ system, ${\cal S}$, and the $D$-wave amplitudes, 
${\cal D}$, ${\cal D}'$ and ${\cal D}''$, are the sums over possible isospin states: 
\begin{align}
&
{\cal M}  = \sum_I {\cal M}_I
\ \ \ \text{for}\ \ {\cal M} = {\cal S},\ {\cal D},\ {\cal D}',\ {\cal D}'',
\end{align}
where $I = 0$ for $\pi^+ \pi^-$, $I = 0,\, 1$ for $K^+ K^-$, and $I = 1$ for $\eta \pi^0$.
Details are given in~\cite{Ali:2010pq}.

\begin{figure}[tb]
\includegraphics[width=0.45\textwidth]{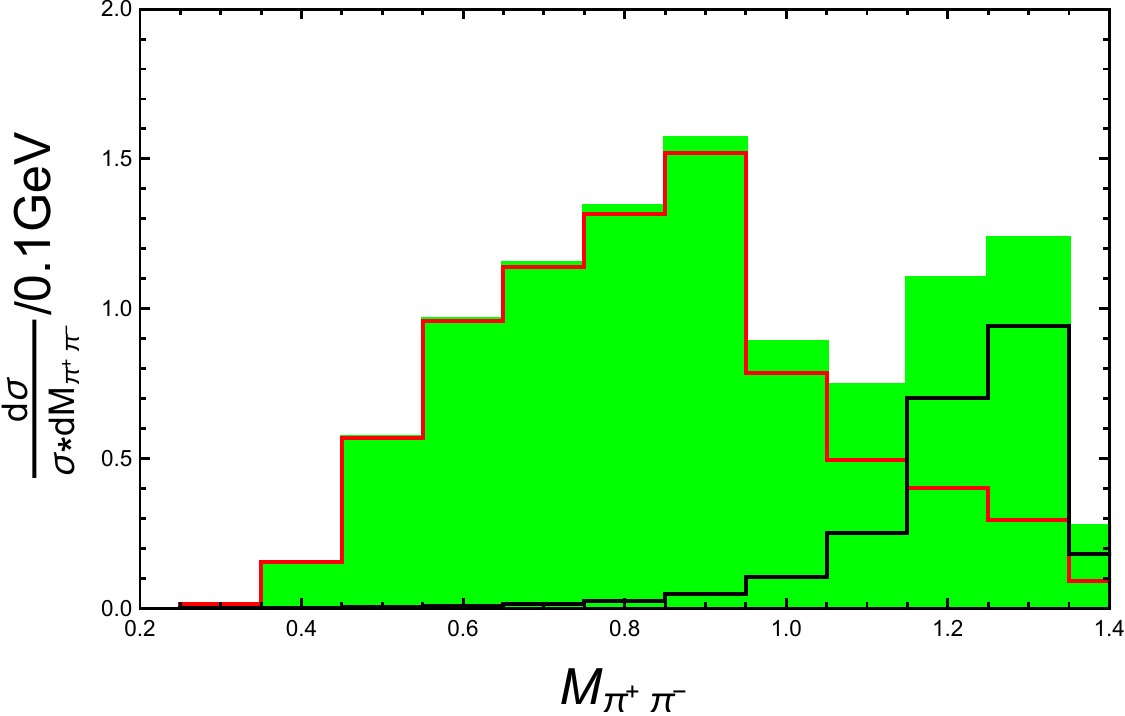}
\includegraphics[width=0.45\textwidth]{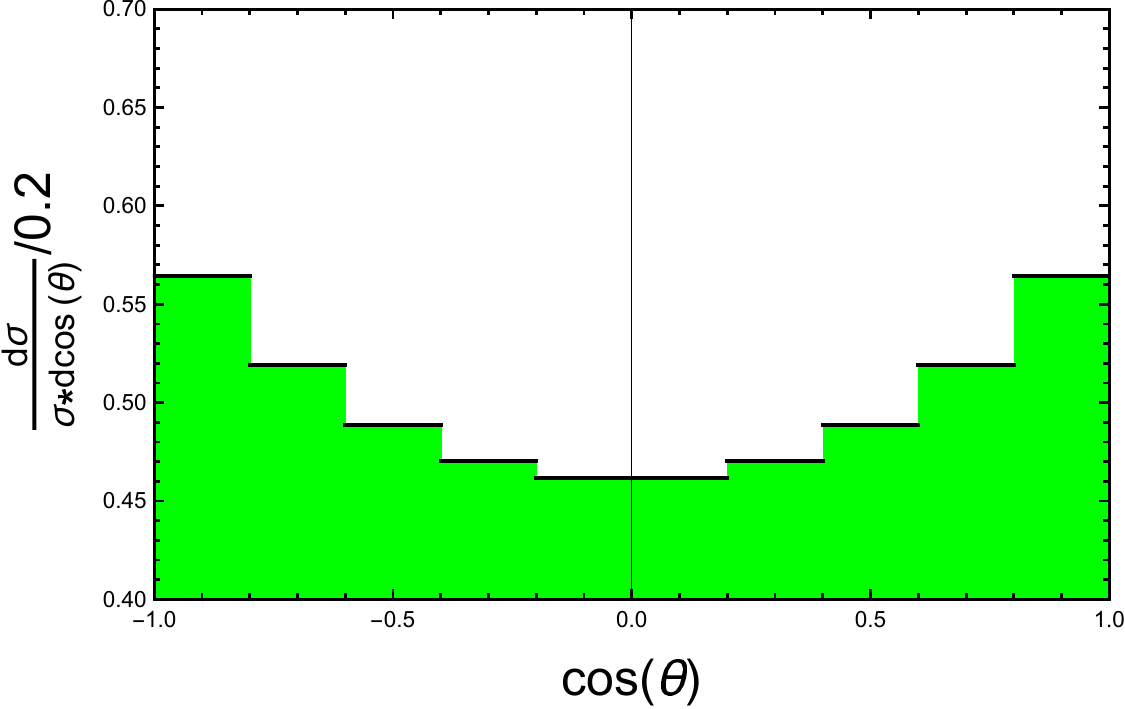}
\caption{\footnotesize 
The normalized resonant $m_{\pi^+ \pi^-}$~(upper plot) and $\cos\theta$~(lower plot) 
distributions for $e^+ e^- \to Y_b (10750) \to \Upsilon(1S) \pi^+ \pi^-$ are shown 
using the coupling constants obtained in~\cite{Ali:2010pq} (green histogram). 
The contributions from $f_0 (500)$ and $f_0 (980)$ scalars (left red curve)  
and $f_2 (1270)$ (right black curve) are indicated in the upper plot.
} 
\label{fig:spectra}
\end{figure}

We concentrate on the process $Y_b (10750) \to \Upsilon (1S)\, \pi^+\pi^-$, in which
the $\sigma = f_0 (500)$, $f_0 (980)$, and $f_2 (1270)$ resonances contribute.  
The $I = 0$ amplitudes are given by the combinations of the resonance amplitudes, 
${\cal M}_0^S$ and ${\cal M}_0^{f_2}$, and the non-resonating continuum amplitudes, 
${\cal M}_0^{1C}$ and ${\cal M}_0^{2C}$: 
\begin{align}
&{\cal S}_0 =
{\cal M}_0^{1C} + (k_1 \cdot k_2) \sum_S {\cal M}_0^S ,
\ \ \
{\cal D}_0 = |\boldmath{k}|^2 {\cal M}_0^{f_2} ,
\nonumber\\
&{\cal D}_0' = {\cal M}_0^{2C} - {\cal D}_0 ,
\ \ \
{\cal D}_0'' =
{\cal M}_0^{2C} + \frac{2 q_0 p_0}{(q \cdot p)}\, {\cal D}_0 ,
\label{eq:izerodef}
\end{align}
where~$S$ runs over possible $I = 0$ scalar resonances, and 
$|\boldmath{k}|$ is the magnitude of the $\pi^+$-meson three momentum in the $\pi^+\pi^-$ rest frame.
The $m_{\pi^+\pi^-}$ and $\cos\theta$ distributions for $e^+ e^- \to Y_b \to \Upsilon (1S)\, \pi^+\pi^-$, 
normalized by the measured cross section $\sigma_{\Upsilon (1S)\, \pi^+\pi^-}^{\rm Belle} = (1.61 \pm 0.16)$~pb 
of the older Belle data~\cite{Abe:2007tk} were fitted in~\cite{Ali:2010pq}, which determined various coupling 
constants. Since these distributions are not available for the new Belle data~\cite{Abdesselam:2019gth}, 
we show in Fig.~\ref{fig:spectra} only the resonant contributions, using the relevant input parameters 
from~\cite{Ali:2010pq}. This illustrates the anticipated spectral shapes, which will be modified 
in detail as the non-resonant contribution is included. The fit can only be undertaken as the experimental 
measurements become available.

The products $\Gamma_{ee} \times {\cal B}$ are measured by Belle, and we take 
$\Gamma_{ee} [Y_b (10750)]$ from~\eqref{mixval}. The corresponding ranges are 
$(0.9 - 3.4)\%$ for ${\cal B}_{\Upsilon(1S) \pi^+ \pi^-}$,  
$(3.9 - 8.9)\%$ for ${\cal B}_{\Upsilon(2S) \pi^+ \pi^-}$, and 
$(1.5 - 1.9)\%$ for ${\cal B}_{\Upsilon(3S) \pi^+ \pi^-}$. 
They are in a reasonable range for the Zweig-allowed decays. We also note that 
due to the dominant tetraquark nature of $Y_b(10750)$, and its quark content, 
decays $Y_b(10750) \to B_s^{(*)} \bar{B}_s^{(*)}$ are not anticipated, in agreement 
with the Belle data~\cite{Abdesselam:2016tbc}.


\vspace*{\baselineskip}
\leftline{\bf 6. Summary}
\vspace*{\baselineskip}


In this work,
we have presented a tetraquark-based interpretation of the Belle data on the new structure $Y_b (10750)$ 
in $e^+ e^-$ annihilation, invoking   the tetraquark-$b \bar b$ mixing  anticipated in the large-$N_c$ limit.
The $b \bar b$-component is used to predict the hadroproduction and Drell-Yan cross sections at the LHC. 
A crucial test of our model is in the $m_{\pi^+ \pi^-}$ and $\cos\theta$ distributions, whose resonant 
contribution is worked out, which is not expected in other dynamical schemes, such as $Y_b (10750)$ 
interpreted as a $D$-wave $b \bar b$ state, with a very large $S-D$ mixing~\cite{Badalian:2009bu}. 
The tetraquark-$Q \bar Q$ mixing scheme suggested here has wider implications.


\vspace*{\baselineskip}
\leftline{\bf Acknowledgements}
\vspace*{\baselineskip}


We thank Changzheng Yuan for informing us of his preliminary results on the electronic width of $Y_b(10750)$ 
and  Satoshi Mishima for his help in checking our code for the distributions shown in Fig.~\ref{fig:spectra} 
and helpful correspondence. The present work was stimulated by discussions at the Workshop on Exotic Hadrons 
held at the  T.\,D.~Lee Institute, Shanghai, June 25--27, 2019, and INPAC, Shanghai Key Laboratory 
for Particle Physics and Cosmology.
The work of W.\,W. is supported in part the National Natural Science Foundation of China 
under Grant Nos. 11575110, 11735010,  11911530088,  and the Natural Science Foundation of Shanghai 
under Grant No. 15DZ2272100. 
A.\,P. and W.\,W. acknowledge financial support by the Russian Foundation for Basic Research 
and National Natural Science Foundation of China according to the joint research project 
(Nos. 19-52-53041 and 1181101282).
This research is partially supported by the ``YSU Initiative Scientific Research Activity'' 
(Project No.~AAAA-A16-116070610023-3). 


\end{document}